\pdfoutput=1
\documentclass[usenatbib]{mn2e}
\usepackage{apjfonts}
\usepackage{amssymb}
\usepackage{amsmath}
\usepackage{ctable}
\usepackage{url}
\usepackage{fixltx2e} 
\usepackage[implicit=false,breaklinks,colorlinks,citecolor=blue]{hyperref}
\usepackage{xcolor}

\newcommand{\apjl}{ApJL}
\newcommand{\apjs}{ApJS}

\newcommand{\aap}{A\&A}

\newcommand{\acknowledgments}{\begin{small}\section*{Acknowledgments}\end{small}}
\newcommand\altaffilmark[1]{$^{#1}$}
\newcommand\altaffiltext[1]{$^{#1}$}
\newcommand{\gizmourl}{\href{http://www.tapir.caltech.edu/~phopkins/Site/GIZMO.html}{\url{http://www.tapir.caltech.edu/~phopkins/Site/GIZMO.html}}}

\title[Passive-Scalar Diffusion in Supersonic Turbulence]{Scaling Laws of Passive-Scalar Diffusion in the Interstellar Medium
\vspace{-0.5cm}}

\vspace{-0.2cm}
\author[Colbrook et al.]{
\parbox[t]{\textwidth}{ 
Matthew J.\ Colbrook,\thanks{E-mail:mjc249@cam.ac.uk}\altaffilmark{1}
Xiangcheng Ma,\altaffilmark{2}
Philip F.\ Hopkins,\altaffilmark{2}
Jonathan Squire\altaffilmark{2,3}} 
\vspace*{6pt} \\
\altaffiltext{1}{Department of Applied Mathematics and Theoretical Physics, University of Cambridge,
Cambridge, CB3 0WA, UK} \\
\altaffiltext{2}{TAPIR, Mailcode 350-17, California Institute of Technology, Pasadena, CA 91125, USA} \\
\altaffiltext{3}{Walter Burke Institute for Theoretical Physics, Pasadena, CA 91125, USA}
\vspace{-0.5cm}
}

\date{Submitted to MNRAS, Accepted 2017 January 27 . Received 2017 January 24 ; in original form 2016 October 18\vspace{-0.6cm}}
\begin{document}
\maketitle

\label{firstpage}

\vspace{-0.2cm}
\begin{abstract}
Passive scalar mixing (metals, molecules, etc.) in the turbulent interstellar medium (ISM) is critical for abundance patterns of stars and clusters, galaxy and star formation, and cooling from the circumgalactic medium. However, the fundamental scaling laws remain poorly understood in the highly supersonic, magnetized, shearing regime relevant for the ISM. We therefore study the full scaling laws governing passive-scalar transport in idealized simulations of supersonic turbulence. Using simple phenomenological arguments for the variation of diffusivity with scale based on Richardson diffusion, we propose a simple fractional diffusion equation to describe the turbulent advection of an initial passive scalar distribution. These predictions agree well with the measurements from simulations, and vary with turbulent Mach number in the expected manner, remaining valid even in the presence of a large-scale shear flow (e.g.\ rotation in a galactic disk). The evolution of the scalar distribution is not the same as obtained using simple, constant ``effective diffusivity'' as in Smagorinsky models, because the scale-dependence of turbulent transport means an initially Gaussian distribution quickly develops highly non-Gaussian tails. We also emphasize that these are mean scalings that only apply to {\em ensemble} behaviors (assuming many different, random scalar injection sites): individual Lagrangian ``patches'' remain coherent (poorly-mixed) and simply advect for a large number of turbulent flow-crossing times.
\end{abstract}

\begin{keywords}
diffusion -- ISM: evolution -- methods: numerical -- methods: analytical -- stars: formation -- galaxies: formation\vspace{-0.5cm}
\end{keywords}

\vspace{-0.5cm}
\section{Introduction}
\label{intro}

Understanding transport processes in the interstellar medium (ISM) is crucial in the study of galaxy evolution, star formation and a wide range of observations in astronomy. For instance, observations of metal abundances in stars give us a window into the past history of galaxies such as our own Milky Way \citep*{2012ARA&A..50..251I}, as well as mapping transitional epochs in the Universe such as the shift from Population III to Population II stars \citep*{2003ApJ...589...35S}. In turn, these provide  clues 
for how to formulate models for stellar enrichment and nucleosynthesis, star and star cluster formation, and even planet formation \citep{2004ApJ...613..898T}.  However, because the ISM is turbulent, metals may mix on small spatial scales relatively easily -- totally independent of how they are transported by bulk flows (e.g.\ galaxy inflows, mergers, outflows) or their original injection (via SNe or other stellar mass-loss processes). Such mixing may alter the interpretation of  observations dramatically. 

To first approximation, individual heavy-element species in the ISM can be treated as passive scalars (although they do participate in dynamics indirectly via cooling). Although passive scalar mixing in subsonic turbulence is well studied in the fluid dynamics community, turbulence in the ISM is highly supersonic (due to efficient radiative cooling) and magnetized. Further, the very large Reynolds numbers of $\sim 10^{10}$ or more \citep{2003ApJ...584..190F} are impossible to simulate directly.  As such, it is important to understand some of the similarities and differences between mixing in neutral incompressible fluids, which tend to follow intuition based on terrestrial flows, and mixing in the supersonic magnetohydrodynamic (MHD) flows that are prevalent in the ISM. 
In this vein, \citet{2010ApJ...721.1765P} have extended the subsonic Obukohov-Corrsin cascade phenomenology \citep{2000Natur.405..639S} to the compressible regime, showing that  mixing time-scales are similar to the time-scales of kinetic energy dissipation, supporting the picture of a cascade of scalar fluctuations in supersonic turbulence.
 Other studies \citep[e.g.][]{2003PhRvE..67d6311K} have focused on a mixing length description  and these ideas have had some success in simple diffusion models  \citep[e.g.][]{2012ApJ...758...48Y}. From such studies, it is clear that the mixing of metal tracers depends on the statistics of turbulence (which depend on parameters such as Mach number) and on the scale considered (in comparison to the physics driving the turbulence).
 

As such, an understanding of mixing in the ISM requires understanding
the statistics of the supersonically turbulent velocity field. These can differ significantly from the velocity statistics in incompressible turbulence due to the formation of shocks, and ``basic'' properties such as the turbulent velocity scaling  remain controversial. Due to this complexity, numerical simulations are key for testing ideas and simple phenomenological arguments. Recently, universal scaling laws for the mass-weighted turbulent velocity have been proposed \citep{2007ApJ...665..416K,2007AIPC..932..393K} and tested in a number of numerical studies \citep{2007AIPC..932..421K,2007AAS...21113803K,2008PhRvL.101s4505S,2010A&A...512A..81F,2010MNRAS.406.1659P,2010PhLA..374.1039S}. Some analytic scaling relations for velocity have also been proposed \citep{2010JFM...644..465F,2011PhRvL.107m4501G,2012JFM...713..482W,2013PhRvE..87a3019B}. These are discussed in \citet*{2013JFM...729R...1K} with an analysis analogous to the Kolmogorov picture of an energy cascade. As well as being important for mixing, these scalings are fundamental inputs to modern theories of star formation via ``turbulent fragmentation'' \citep*{2009PhRvL.102c4501P,2013MNRAS.430.1653H}.

Due to the very high Reynolds numbers, most studies are forced to adopt a Subgrid-Scale Model (SGS) to simulate ISM mixing, since it is not possible to resolve the viscous scale
(but see, for example,  \citealt{2015MNRAS.449.2588P}). A popular example is the \citet{1963MWRv...91...99S} model, which adopts a locally-constant eddy diffusivity proportional to the resolved strain rate tensor. \citet*{2010MNRAS.407.1581S} found that such subgrid feedback models alter the metal enrichment in smoothed-particle hydrodynamics (SPH) simulations significantly. However, this model fails to describe the scale-dependence of turbulence; moreover, it was derived for highly subsonic, nonmagnetized turbulent flows, without bulk velocity shear -- none of these assumptions hold in the ISM. As highlighted by \citet*{2008MNRAS.387..427W}, using these simple scalings without a more physically-motivated formulation of dissipation can lead to order-of-magnitude errors, and even their qualitative ``correctness'' and convergence may not be well defined.

In this paper we seek to study the local diffusion properties of a passive tracer initially ``injected'' into a supersonically turbulent medium. Following the above discussion, we investigate the possibility of a scaling law for the diffusivity that is dependent on wavenumber/length scale, in a similar manner to \emph{Richardson diffusion} in subsonic flows \citep{Richardson709}.  The results can be viewed as an extension of \citet{2003PhRvE..67d6311K} to all scales in the turbulence, or of Richardson diffusion into the highly-compressible supersonic regime. While there have been arguments for describing anomalous diffusion in this way in other physical situations \citep{2010ASSP...20...63S,metzler2000random,balescu1995anomalous,balakrishnan1985anomalous}, and in flux freezing in subsonic MHD turbulence \citep{eyink2013flux}, so far as we know this is the first time such a model has been tested against numerical data for supersonic  turbulent transport in the ISM . We shall argue that Richardson's scaling for a diffusivity $D(l)\propto{}\epsilon^{1/3}l^{4/3}$, where $\epsilon$ is the mean energy dissipation, becomes steeper in supersonic flows due to the different scaling of the velocity structure functions. More concretely, we argue that within an inertial range of wavenumbers, the process can be described by a \emph{fractional diffusion} process, with ${\partial_t}\widehat{\theta}(\boldsymbol{\bf k})=-{|\boldsymbol{\bf k}|^2}{\kappa(\boldsymbol{\bf k})}\widehat{{\theta}}(\boldsymbol{\bf k})$ with $\kappa\propto{}\mathcal{M}|\boldsymbol{\bf k}|^{-\alpha}$ and $\alpha \sim 1+\zeta(1)$. Here $\theta$ denotes metal density, $\mathcal{M}$ Mach number and $\widehat{x}$ denotes the Fourier transform of $x$. Within uncertainties, the model agrees well with numerical simulations of isothermal turbulence (we ignore the details of radiative cooling and heating, as is common in ISM turbulence studies). We consider  3D neutral-fluid turbulence with a turbulent Mach number $\mathcal{M}\approx 7$, and then various extensions to test the robustness of the theory:  magnetohydrodynamics at various Mach numbers and in two and three dimensions, and the presence of a mean shear flow (simulating rotation of a galactic disc). We ignore self-gravity because we do not wish to explicitly follow star formation.

The paper is organized as follows: in \S~2, we outline the theory of classical mixing length descriptions and our argument; \S~3 describes our method and simulations; results and comparison with theory are discussed in \S~4; our conclusions and the implications for metal transport in the ISM are outlined in \S~5.

\vspace{-0.5cm}
\section{Theory}
\label{theory1}

We do not give a full account of the statistics of mixing in supersonic turbulence and refer the reader to papers such as \citet{2010ApJ...721.1765P} for such a treatment where the classic picture of a cascade of scalar fluctuations is applied to the supersonic regime. Crucially, in agreement with our argument below, it was found there (and in subsequent studies) that, for a wide range of Mach numbers, the mixing time-scale was proportional to the turnover time of eddies at the length scale of the scalar sources. Here we consider the simple case where a tracer is ``released'' in an initially highly-concentrated ($\delta$-function or ``point source'') distribution around an injection site, and attempt to follow its ``diffusion.''

\citet{Taylor01011922} introduced the formula
\begin{eqnarray}
\frac{\mathrm{d}}{\mathrm{d}t}\xi=\int\limits_{0}^t \ \left\langle \boldsymbol{\bf v}(\boldsymbol{\bf x}(0),0)\boldsymbol{\cdot}\boldsymbol{\bf v}(\boldsymbol{\bf x}(t'),t')\right\rangle\mathrm{d}t'
\label{Taylor}
\end{eqnarray}
where $\xi=\langle{|\boldsymbol{\bf x}(t)-\boldsymbol{\bf x}(0)|}^2\rangle$  is the ensemble average of particle displacements following a Lagrangian trajectory and $\boldsymbol{\bf v}$ denotes an Eulerian velocity. Conceptually, this links the statistics of the Lagrangian and Eulerian viewpoints. In the above, we assume isotropy so that $\xi$ need not be defined for different directions (this does not have to be true in MHD turbulence, but we will show below it is valid in an {\em ensemble} sense). For times much larger than the autocorrelation time, $\tau$, we expect the right-hand side to be a constant and hence the left-hand side gives a definition of a Lagrangian diffusivity, $D={{\mathrm{d}}\xi}/{\mathrm{d}t}$. 

If $f(\boldsymbol{\bf r},t)$ denotes the probability distribution for finding a particle (or element of tracer) at position ${\bf r} \equiv {\bf x}(t) - {\bf x}(0)$ after a time $t$ then, {\em if we assume} the position has a Gaussian distribution \cite[corresponding to a first-order Markov process or random walk, see][]{sawford2001turbulent}, we expect \citep[following][]{1949AuSRA...2..437B} a diffusion equation to hold:
\begin{eqnarray}
{\partial_t}{f}=D{\nabla^2}{f}.
\label{Batchelor}
\end{eqnarray}
In a real turbulent flow in the inertial range, the assumption of Gaussian statistics is far from correct. If we consider turbulence as a hierarchy of eddies, we can attach to each eddy a length scale $\hat{l}$ and a velocity scale $\hat{v}$. These determine the eddy turnover time as $\hat{\tau}={\hat{l}}/{\hat{v}}$. For $t<\hat{\tau}$ individual elements (molecules, metal species, etc.) which are ``injected together'' are strongly correlated, which leads to the estimate $|\mathbf{x}(t)-\mathbf{x}(0)|\approx{\hat{v}}\,t$ and $D(t)\approx{}2\,{{\hat{v}}^2}\,t$. This is simply the ballistic motion -- pure advection at a locally constant velocity -- of a tracer in the eddy's local flow. For $t\gg\hat{\tau}$ the eddy has dispersed and destroyed the correlation in the velocity field, implying we should replace $t$ by $\hat{\tau}={\hat{l}}/{\hat{v}}$, and giving the estimate $D(t)\approx{}2{\hat{l}}{\hat{v}}$ for the diffusion coefficient. Hence the expected scalings are
\begin{eqnarray}
D(t)\approx
\begin{cases}
2\,{{\hat{v}}^2}\,t,& t<\hat{\tau}\\
2\,{\hat{l}}\,{\hat{v}},& t\gg\hat{\tau}
\end{cases}
\label{Dscale}
\end{eqnarray}
For our purposes, note the shift of scale dependence on $\hat{v}$ and the scale dependence of the diffusion constant for $t\gg\hat{\tau}$. \citet{2003PhRvE..67d6311K} found that this approach can be continued into the compressible regime by introducing a shock travel length ${l}^*$ and rms velocity $v^*$, leading to the crossing time ${\tau}^*={l^*}{v^*}$. They used $l^*=L/{k_f}$, where $L$ is the size of the region under consideration and $k_f$ the forcing wavenumber for their numerical simulations. We now extend these ideas by studying the analogous scalings of $D$ for a range of wavenumbers, not just those which contain the most energy.

Similar arguments to those described in the previous paragraph were originally proposed by \citet{Richardson709}, who suggested that dispersion of nearby particles (two-point statistics) is diffusive with $D\sim{}r^{4/3}$. This is usually stated in the form $\langle r^2(t)\rangle\sim\epsilon{}\,t^{3}$ (here, $\langle r^2(t)\rangle$ denotes the mean square particle separation and $\epsilon$ the energy dissipation rate). The physical argument is that, in the inertial range, only eddies with a scale similar to the particle separation act to increase the separation. For, say, a ``patch'' or concentration of scalar density with some physical scale, much smaller eddies simply stir scalars {\em within} the patch, while much larger eddies simply advect the entire patch. As the separation (``patch size'') increases, it encounters ``resonant'' eddies of increasing size and hence diffuses faster. The scaling of eddies with physical scale is quantified in terms of structure functions \citep[e.g.][]{1975mit..bookR....M}, which are defined by 
\begin{eqnarray}
{S_p}(l)\equiv \langle |v_l|^p \rangle=\left\langle{{\left|{\boldsymbol{\bf v}}({\boldsymbol{\bf x}}+{\boldsymbol{\bf l}})-{\boldsymbol{\bf v}}({\boldsymbol{\bf x}})\right|}^p}\right\rangle,
\label{structure}
\end{eqnarray}where $\langle \cdot \rangle$ is the ensemble average over all directions $\boldsymbol{\bf l}$ with $|{\boldsymbol{\bf l}}|=l$, and $v_l$ is the velocity difference over scale $l$ (we assume isotropic turbulence). 
For large Reynolds numbers (i.e., far from the energy injection scale and the dissipation scales) and  above or below the sonic scale where $v_l\sim c_s$, the turbulence is scale free and the structure functions must satisfy a power law scaling ${S_p}(l)\sim{l^{\zeta(p)}}$.

The standard subsonic Kolmogorov scaling gives $\zeta(p)\approx{}p/3,$ which has been supported with experimental data for small $p$. If we take the regime where $D\sim{\hat{l}}{\hat{v}}$ and $\left\langle v_l \right\rangle\sim l^{\zeta(1)}$ then we obtain $D\sim{}l^{1+\zeta(1)}\sim{}l^{4/3}$ in agreement with Richardson's scaling. For the regime $t<\hat{\tau}$ we obtain $D\sim{}l^{\zeta(2)}t$ and Kolmogorov gives $\zeta(2)=2/3$. In summary, if we assume for large Reynolds numbers turbulence relaxes into a self-similar state with fluctuations obeying Eq. (\ref{structure}), then upon taking ensemble averages of Eq. (\ref{Dscale}) it is reasonable to expect
\begin{eqnarray}
D\sim
\begin{cases}
{l}^{\zeta(2)}\,t\ & t<\hat{\tau}\\
{l}^{\zeta(1)+1}\ & t\gg\hat{\tau}
\end{cases}.
\label{Dscale2}
\end{eqnarray}
The velocity structure function scalings of supersonic turbulence differ from those in subsonic turbulence. A common estimate in the supersonic cascade at $\mathcal{M}\gg 1$ is  $\zeta(1)\approx{}0.5$, with 
some decrease as $\mathcal{M}$ approaches $1$. The value for $\zeta(2)$ is less well known but should be in the range $\zeta(2)\sim{}0.8-1$. Because these are larger than the subsonic estimate $\zeta(p)\approx p/3$, we  expect a stronger scaling of $D$ with $l$ in supersonic compared  to subsonic turbulence. 

Since the diffusion process described in Eq. (\ref{Dscale2}) explicitly depends on the scale being considered, it is most natural to consider the diffusion of the tracer in the Fourier domain:\footnote{Working in $n$ dimensions we define the Fourier transform of a function $f$ with domain $\Re^n$ as
\begin{eqnarray}
\widehat{f}(\boldsymbol{\bf k})=\int\limits_{\Re^n} f(\boldsymbol{\bf x})\exp(-i\boldsymbol{\bf x}\boldsymbol{\cdot}\boldsymbol{\bf k})\ \mathrm{d}\boldsymbol{\bf x}.
\label{FTconvention1}
\end{eqnarray}}
\begin{eqnarray}
{\partial_t}\widehat{{\theta}}(\boldsymbol{\bf k})=-{k_i}\,{{\kappa_{ij}}(\boldsymbol{\bf k})}\,{k_j}\,\widehat{{\theta}}(\boldsymbol{\bf k}).
\label{equation_of_state}
\end{eqnarray}
Here and throughout, $\theta$ denotes the passive scalar (e.g.\ metal) density profile. In  isotropic turbulence,  we may take $\kappa_{ij}={\kappa}\,\delta_{ij}$ and if we assume a power law scaling (i.e., scale invariance of the turbulence) $\kappa(k)\propto{}k^{-\alpha}$ (with $k\equiv |{\bf k}|$) then Eq. (\ref{equation_of_state}) corresponds to a fractional diffusion equation
\begin{equation}
\partial_t{\theta}=-(-\Delta)^{(2-\alpha)/2}\theta,
\end{equation} where $\Delta$ denotes the Laplacian. Letting $\beta=2-\alpha$, the case $\beta=2$ corresponds to a standard diffusion equation. In the case of $0<\beta<2$ we obtain the evolution equation for the probability distribution function of a stable\footnote{This means that the probability density of any linear combination of uncorrelated random variables with this distribution coincides with the original distribution up to rescaling. } L{\'e}vy flight \citep*{klages2008anomalous}. This extends the standard diffusion model to situations where assumptions of locality, Gaussianity and lack of long-range correlations fail to hold. A similar model has met with some success for the description of transport in plasma turbulence \citep*{del2005nondiffusive}.

Taking $\kappa={}k^{-\alpha}$, the solution of Eq. (\ref{equation_of_state}) with an initial point source is
\begin{eqnarray}
\theta(r,\,t)=\frac{1}{(2\pi)^n}\int\limits_{\Re^n} \exp(i\,\boldsymbol{\bf k}\boldsymbol{\cdot}\boldsymbol{\bf x}-k^{\beta}t)\ \mathrm{d}\boldsymbol{\bf k} \sim \frac{t}{r^{n+\beta}} \sim \frac{t}{r^{2+n-\alpha}}
\label{FTconvention2}
\end{eqnarray}
for large $r = |{\bf r}|$, where $n$ is the dimension of the system. Using this, it is easy to see that the fractional moments $\langle r^{\delta} \rangle$ diverge for $\delta\geq \beta$. However, for $ 0\leq\delta<\beta$ we have
\begin{eqnarray}
\langle r^\delta \rangle \sim t^{\frac{\delta}{\beta}}
\end{eqnarray}
and one can extend this scaling to larger $\delta$ by accounting for cut-off effects \citep{metzler2000random}. Taking $\beta=2-\alpha=1-\zeta(1)$ gives a prediction which agrees with Richardson's scaling $\langle r^2(t)\rangle\sim\epsilon{}\,t^{3}$ (albeit for single-particle separation), while for supersonic turbulence with the estimate $\zeta(1)\approx 1/2$ we obtain $\langle r^2 \rangle \sim t^{4}$. Similarly, for small $k$ we may expect $\kappa\propto{}k^{-\zeta(2)}t$, and so both scalings become steeper as we increase the Mach number. 

The considerations of the previous paragraph suggest that we should be able to {\em measure} a well-defined, scale-dependent ``diffusivity'' by measuring
\begin{eqnarray}
\kappa=-\frac{{\partial_t}\widehat{{\theta}}}{{k^2}\widehat{{\theta}}}
\label{kappadef2}
\end{eqnarray}
from numerical simulations. This is consistent with the classical case with constant diffusion coefficient and has the advantage of being straightforward to compute numerically. If we take the scalings in Eq. (\ref{Dscale2}) then one may conjecture that
\begin{eqnarray}
\kappa=A\mathcal{M}k^{-\alpha}
\label{expected_rel}
\end{eqnarray}
within the inertial range of wavenumbers.

\begin{figure}
\begin{tabular}{l}
\includegraphics[width=0.97\columnwidth,trim={0cm 3.8cm 0cm 2.9cm},clip]{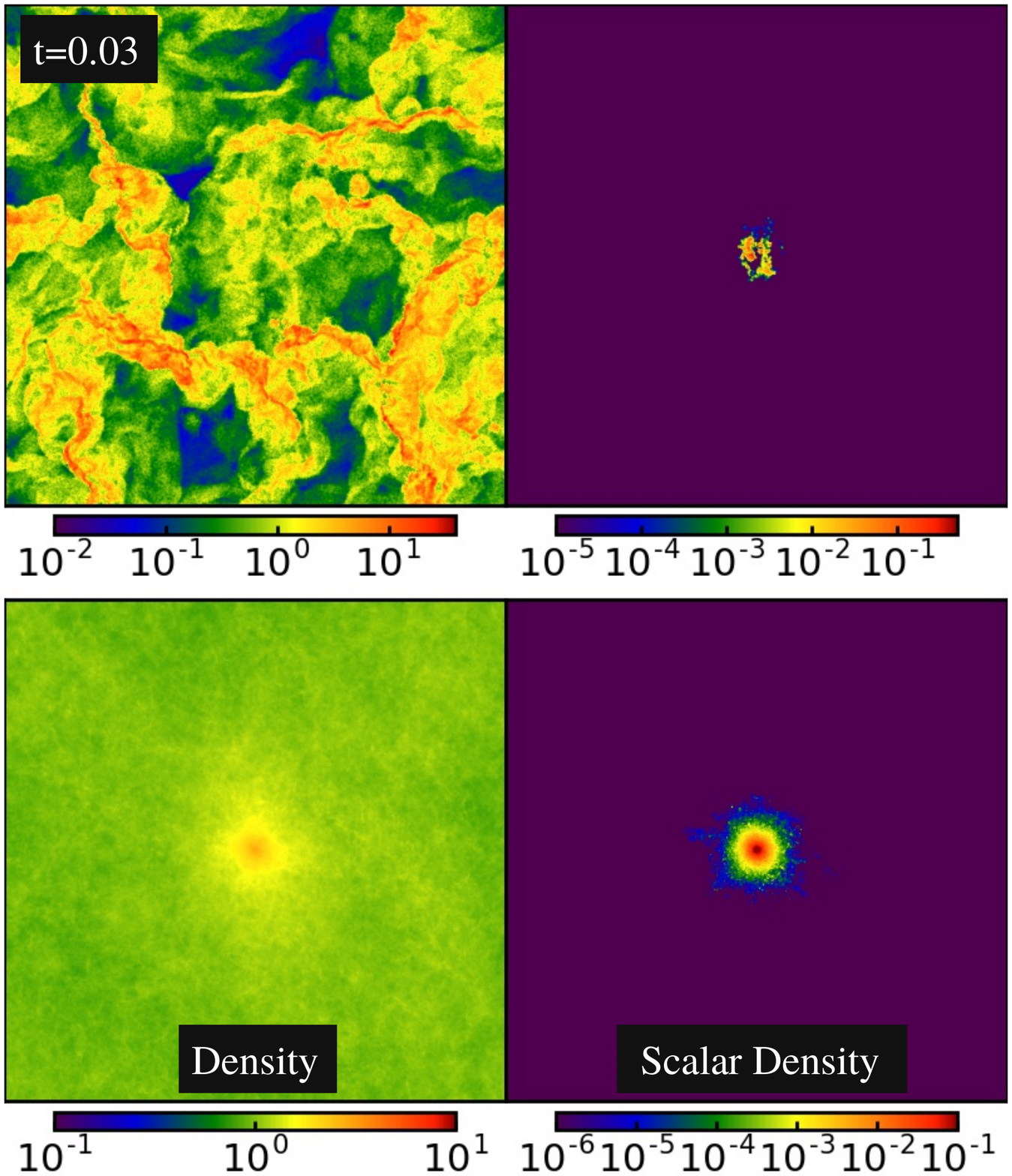} \\
\end{tabular}
    \caption{{\em Top:} Gas density $\rho$ ({\em left}) and passive scalar/metal density $\theta$ ({\em right}), denoted by \textcolor{blue}{colour} (as labeled), in our fiducial run. We show all quantities directly ``as they are'' in the code at a time $t=0.03$ (in code units; a few crossing times at the initial injection scale) after a single point-like ``injection'' of the metals/scalars into the center of the box. Clearly, the distribution even after a few small-scale turbulent crossings is highly anisotropic, dominated by advection and shear along field lines -- it does not resemble diffusion.
    {\em Bottom:} Same, but repeating the ``injection'' process at $80$ different random locations in the box, each at $15$ different random initial times ($1200$ injections in total), then averaging all of the resulting maps together (after re-centering each on the center of the tracer mass distribution). This is the ``averaged'' profile $\theta(r,\,t)$ that we analyze. In {\em ensemble-average}, the distribution is both isotropic and qualitatively diffusion-like.  The re-centered density field shows a maximum at the box centre because we remove the centre-of-mass motion of each tracer packet before averaging over realizations.
    \label{fig:averaging}}
\end{figure}

\begin{figure}
\vspace{-0.3cm}
\includegraphics[width=1\columnwidth]{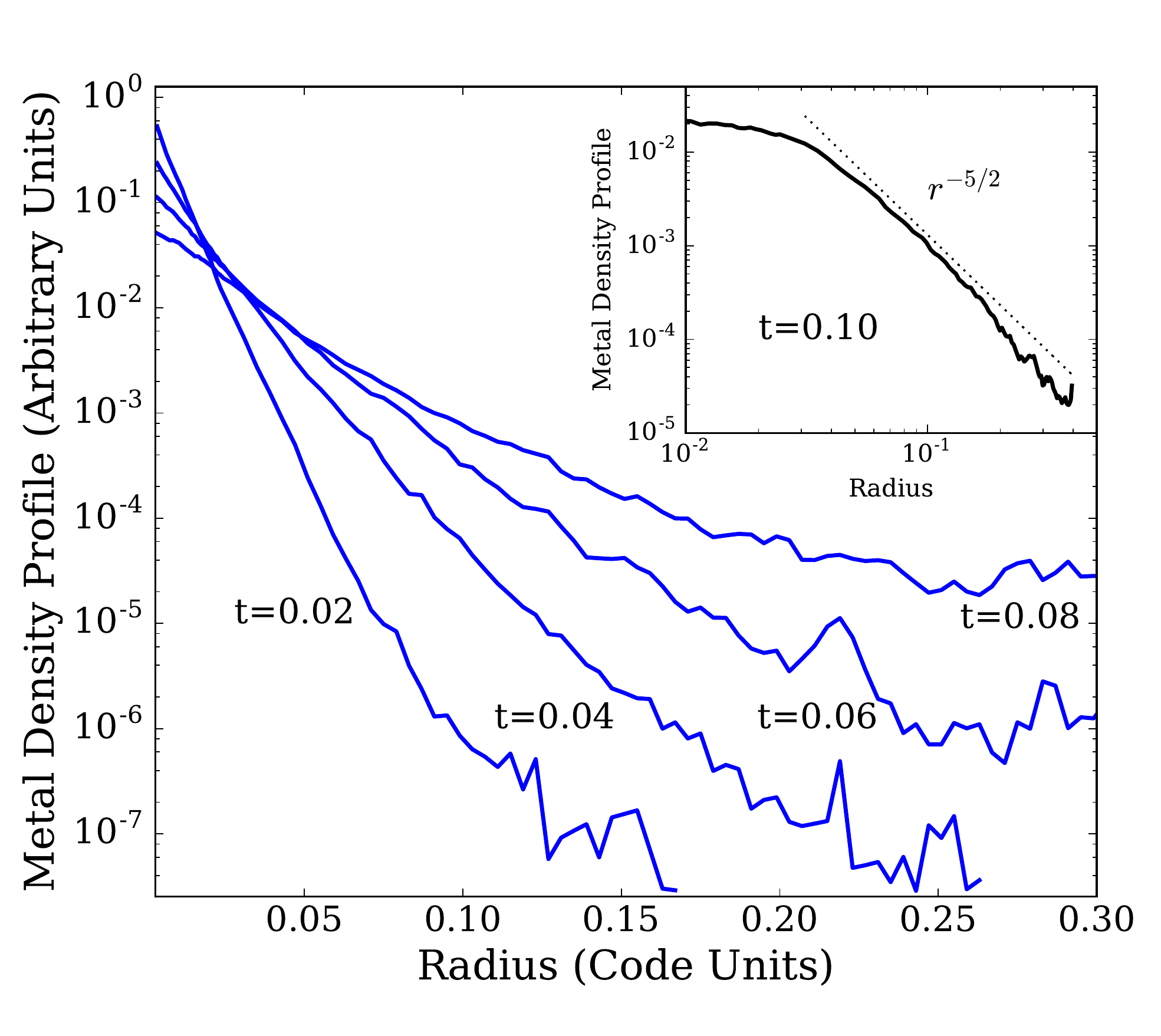} 
    \vspace{-0.5cm}
    \caption{The ensemble-averaged metal density profile $\theta(r,\,t)$ at time $t$ after tracer release in the 2D non-shearing {MHD}  simulation at $\mathcal{M}\approx5$. A scale-independent ``effective diffusivity'' $\kappa$ would produce a Gaussian profile here ($\theta \propto \exp{[-C\,r^{2}]}$; this would appear as a parabola in the main figure), but this is not a good description of the profile at {\em any} time shown, especially in the tails. The inset shows the profile at the final time $t=0.1$ in log-log space, demonstrating that the tails have a clear power-law behavior. We compare the analytically predicted power-law slope (\S~\ref{sec:diffusivity.not.constant}) from our theoretical model of scale-dependent diffusion ($\theta \propto r^{-5/2}$); this agrees well with the simulations. Analogous plots for the other simulations are qualitatively similar, but the higher resolution in this case allows for better identification of the power-law behavior at late times.
    \label{fig:tails}}
\end{figure}

\begin{figure}
\begin{tabular}{l}
\vspace{-0.2cm}
\includegraphics[width=0.99\columnwidth]{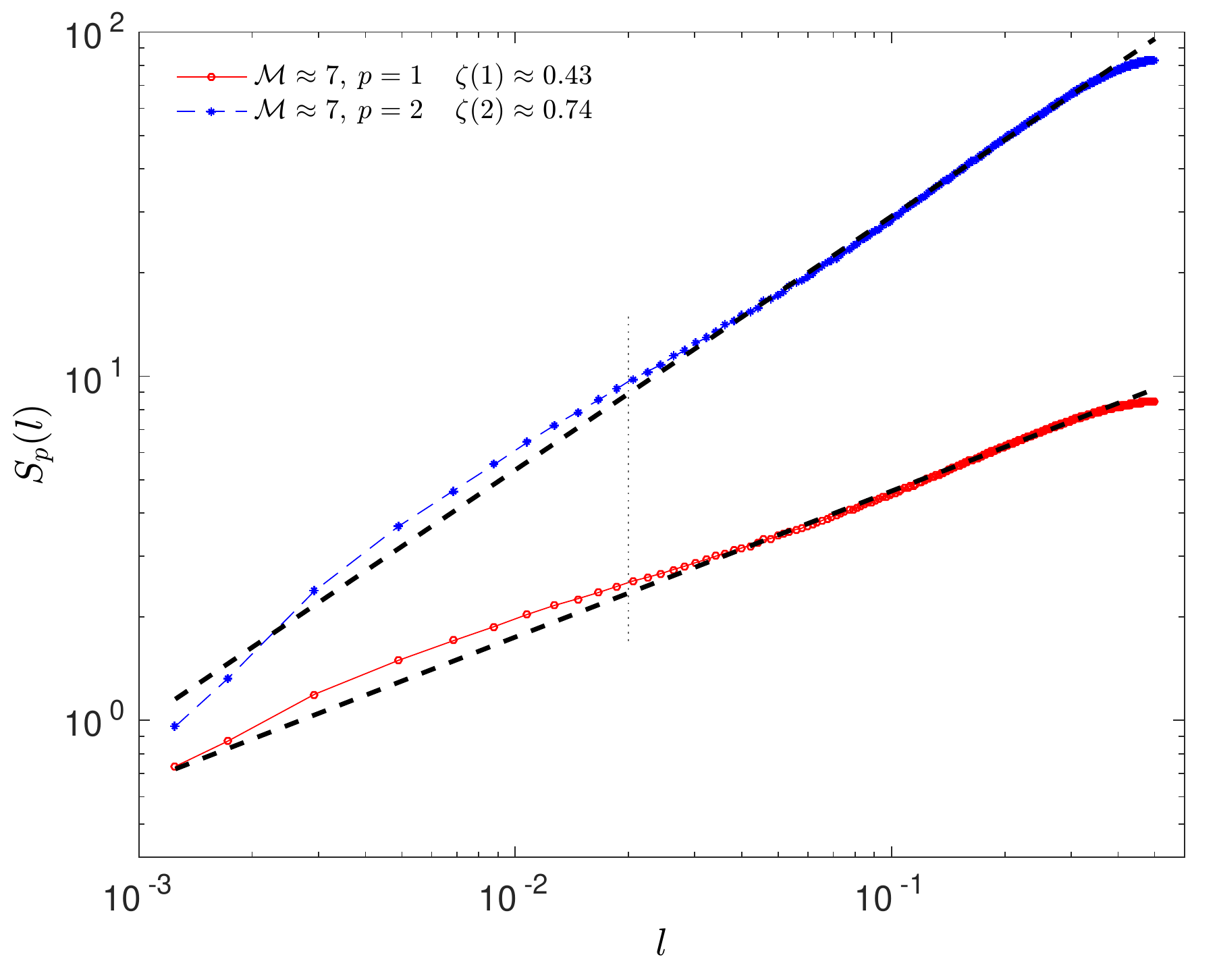} \\
\vspace{-0.3cm}
\includegraphics[width=0.99\columnwidth]{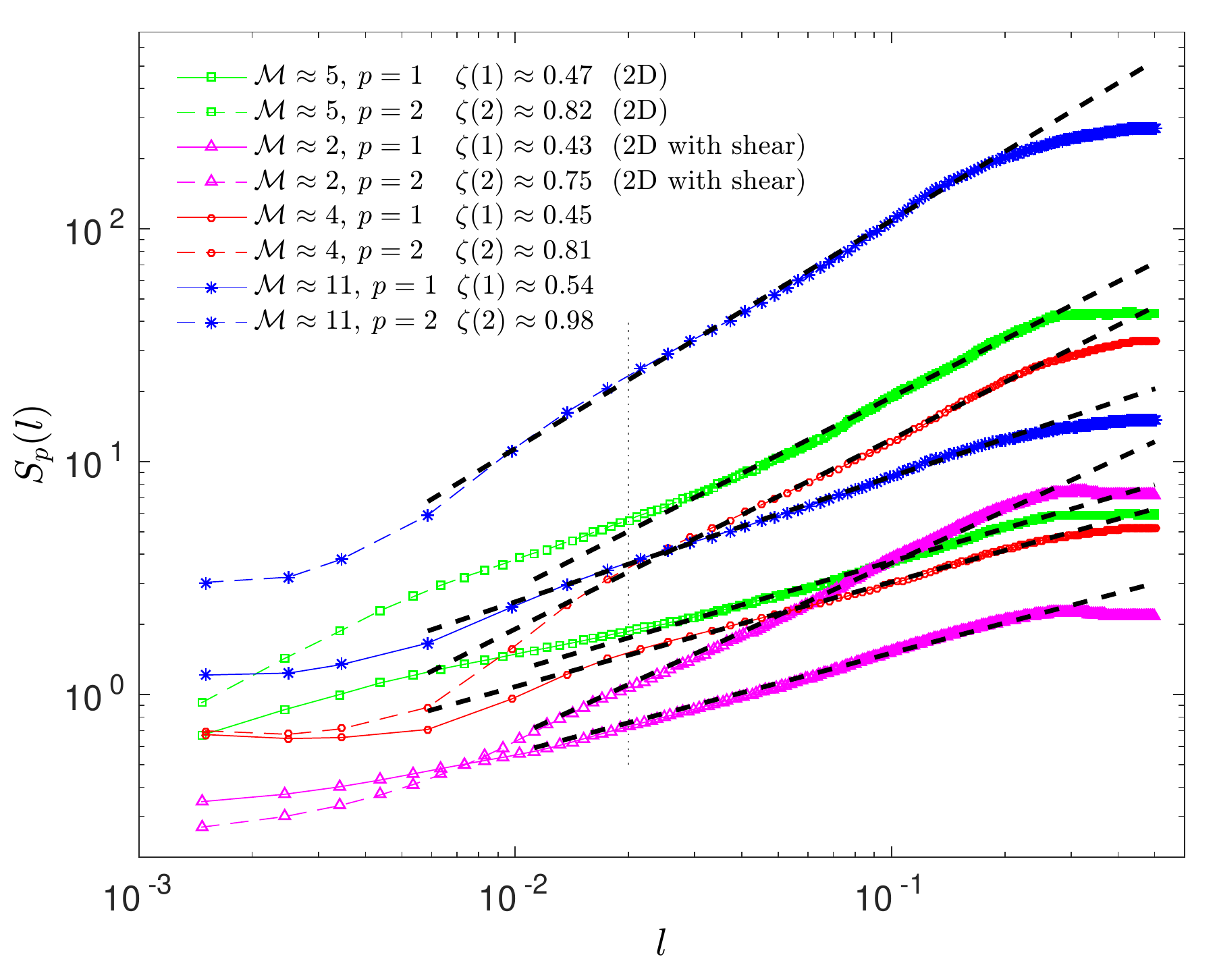} 
\end{tabular}
    
    \caption{{\em Top panel: }Structure functions $S_{1}(l)$ and  $S_{2}(l)$  [see Eq.~\eqref{structure}] for our fiducial simulation at $\mathcal{M}\approx7$. We measure the scaling exponents $\zeta(1)$ and $\zeta(2)$ by fitting a power law between $l_{\mathrm{fit}}\approx 2\times 10^{-2}$ (shown by the  vertical dotted line) and the point where the $S_{1}(l)$ flattens at the largest scales ($l \approx 4 \times 10^{-1}$). This value for $l_{\mathrm{fit}}$ is chosen so as to not include the contribution from subsonic turbulence at smaller $l$, and throughout the fitting region we see that $S_{p}$ is well approximated by a power law.    {\em Bottom panel:} Same as the top panel but for each of the MHD simulations. The scaling exponents $\zeta(p)$ are again measured between $l_{\mathrm{fit}} \approx 2\times 10^{-2}$ and the turnover at large scales, since this approximately captures the power-law range in each case (see Sec.~\ref{sec:Fourier scaling results} for discussion).  The structure function in the 2D shearing simulation is measured in the direction perpendicular to the shear. As expected,  $\zeta(p)$ increases slightly with Mach number, and is similar between 2D and 3D simulations.
    \label{fig:struct}}
\end{figure}

\begin{figure}
\includegraphics[width=0.97\columnwidth,trim={3.1cm 8cm 3.9cm 9.5cm},clip]{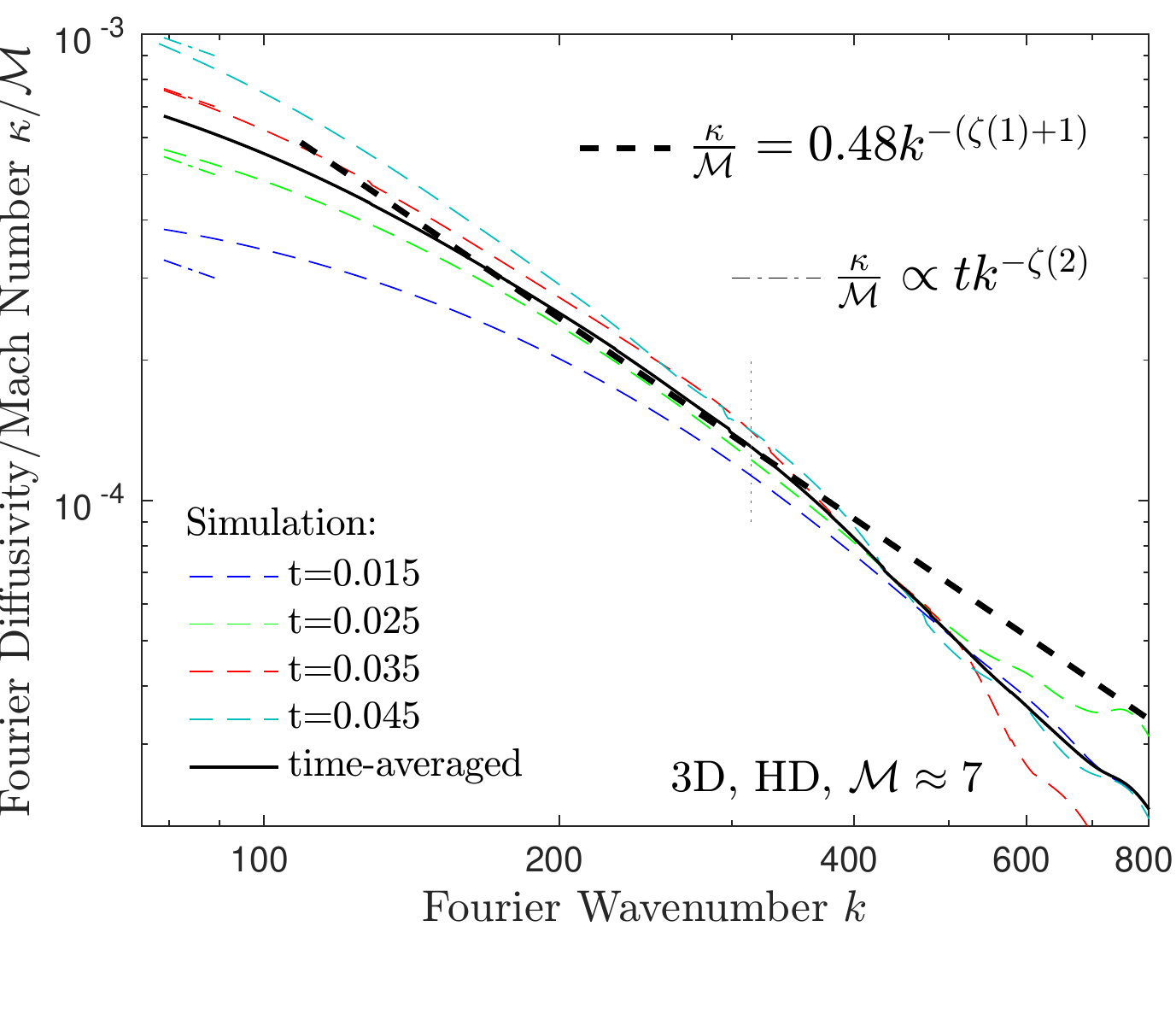}
	\vspace{-0.3cm}
    \caption{{}``Fourier Diffusivity'' $\kappa$ as a function of $k$ (effective diffusivity of modes with wavenumber $k$, defined as $\kappa \equiv -{\partial_t}\widehat{{\theta}}/({k^2}\widehat{{\theta}})$, see $\S$\ref{theory1}), normalized by the box Mach number $\mathcal{M}$. We plot this at different times and the mean over all times sampled (as labeled). We compare our theoretical prediction from Eq.~\ref{Dscale2} and Eq.~\ref{expected_rel}, for both the inertial range where we expect a time-independent scaling $\kappa/\mathcal{M} = A\,k^{-(\zeta(1)+1)}$ (Eq.~\ref{expected_rel}; thick-dashed lines), and for small $k$, where we expect a time-dependent scaling $\kappa/\mathcal{M} \propto t\,k^{-\zeta(2)}$ (Eq.~\ref{Dscale2}; dot-dashed lines). We use the $\zeta$ values directly measured for the same simulation in Fig.~\ref{fig:struct}. The analytic scalings agree relatively well with the simulation. In the simulation's inertial range from Fig.~\ref{fig:struct}, we see an approximately  time-independent scaling with universal constant coefficient $A\approx 0.5$. This corresponds to the simple physical scaling $\kappa \sim v_{t}(l)\,l \sim k^{-1.5}$ in real-space. At small $k$, the scaling is dominated by simple ``ballistic motion'' of Lagrangian gas elements. The dotted line illustrates $k_{\mathrm{fit}}=2\pi/l_{\mathrm{fit}}$, to aid in  comparison with Fig.~\ref{fig:struct}.
    \label{fig:kspaceHD}}
\end{figure}

\begin{figure*}
\begin{tabular}{cc}
\vspace{-0.7cm}
\includegraphics[width=0.97\columnwidth,trim={3.1cm 8cm 3.5cm 9.5cm},clip]{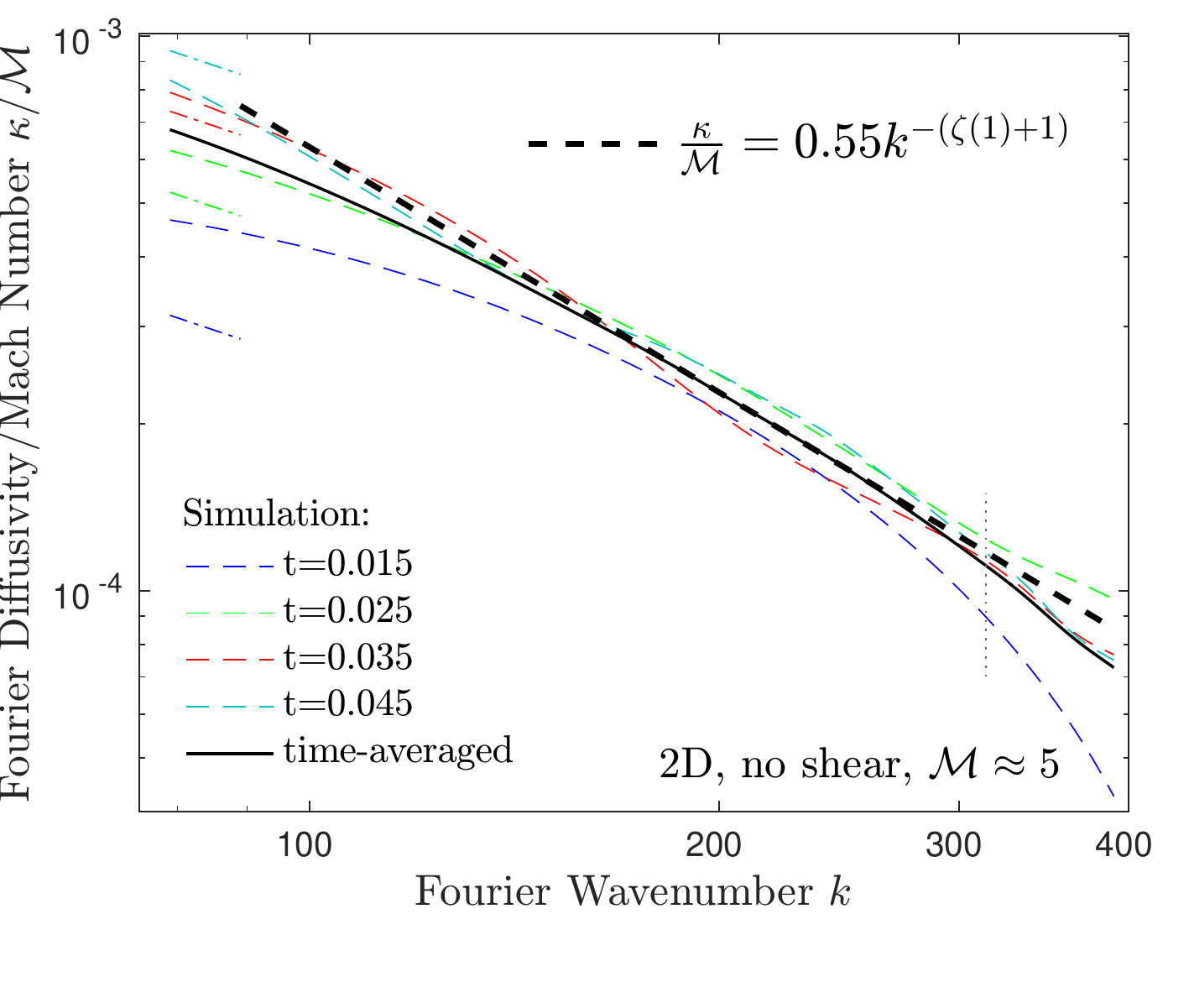} &

\includegraphics[width=0.97\columnwidth,trim={3.1cm 8cm 3.5cm 9.5cm},clip]{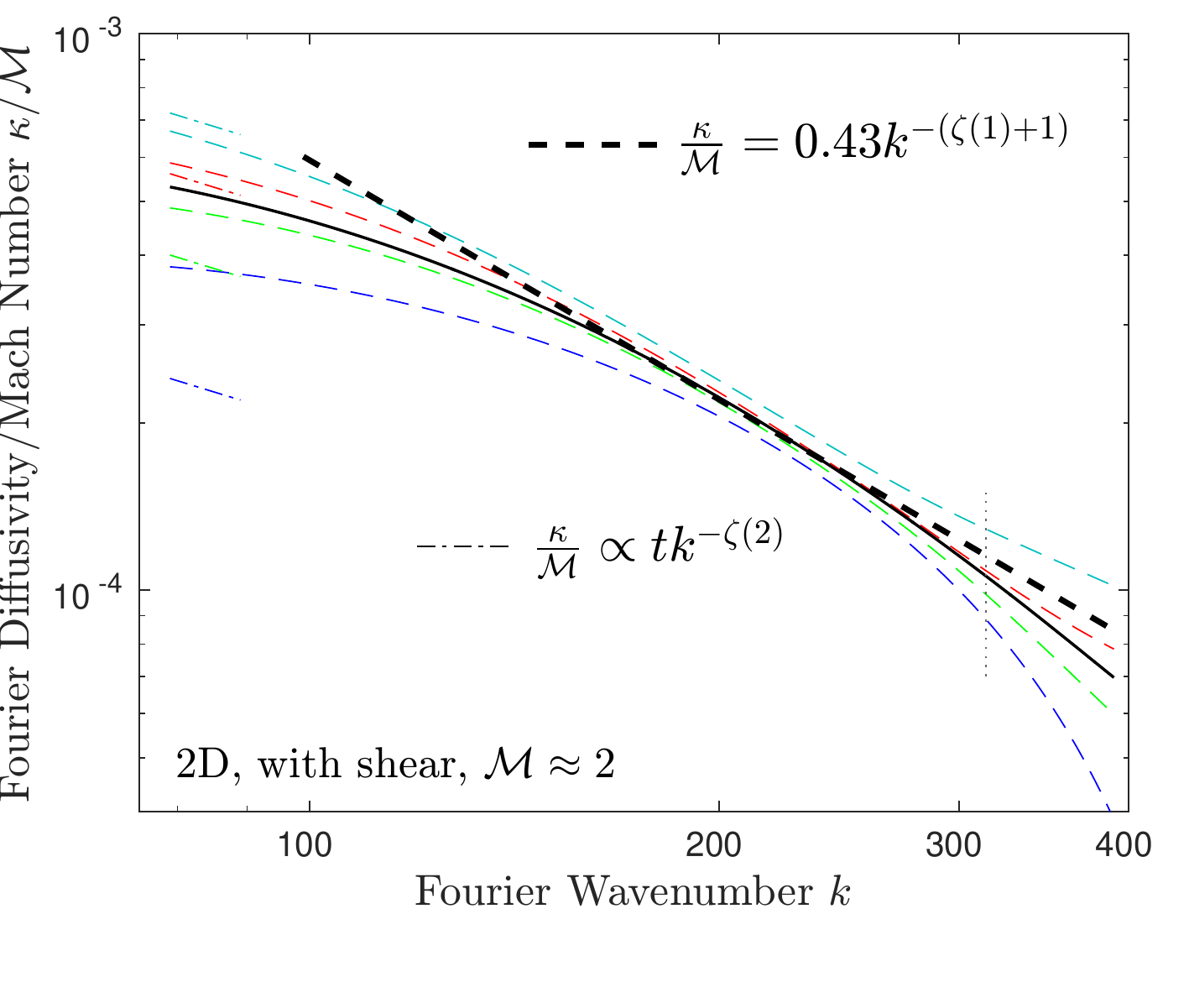} \\
\includegraphics[width=0.97\columnwidth,trim={3.1cm 9cm 3.5cm 9.5cm},clip]{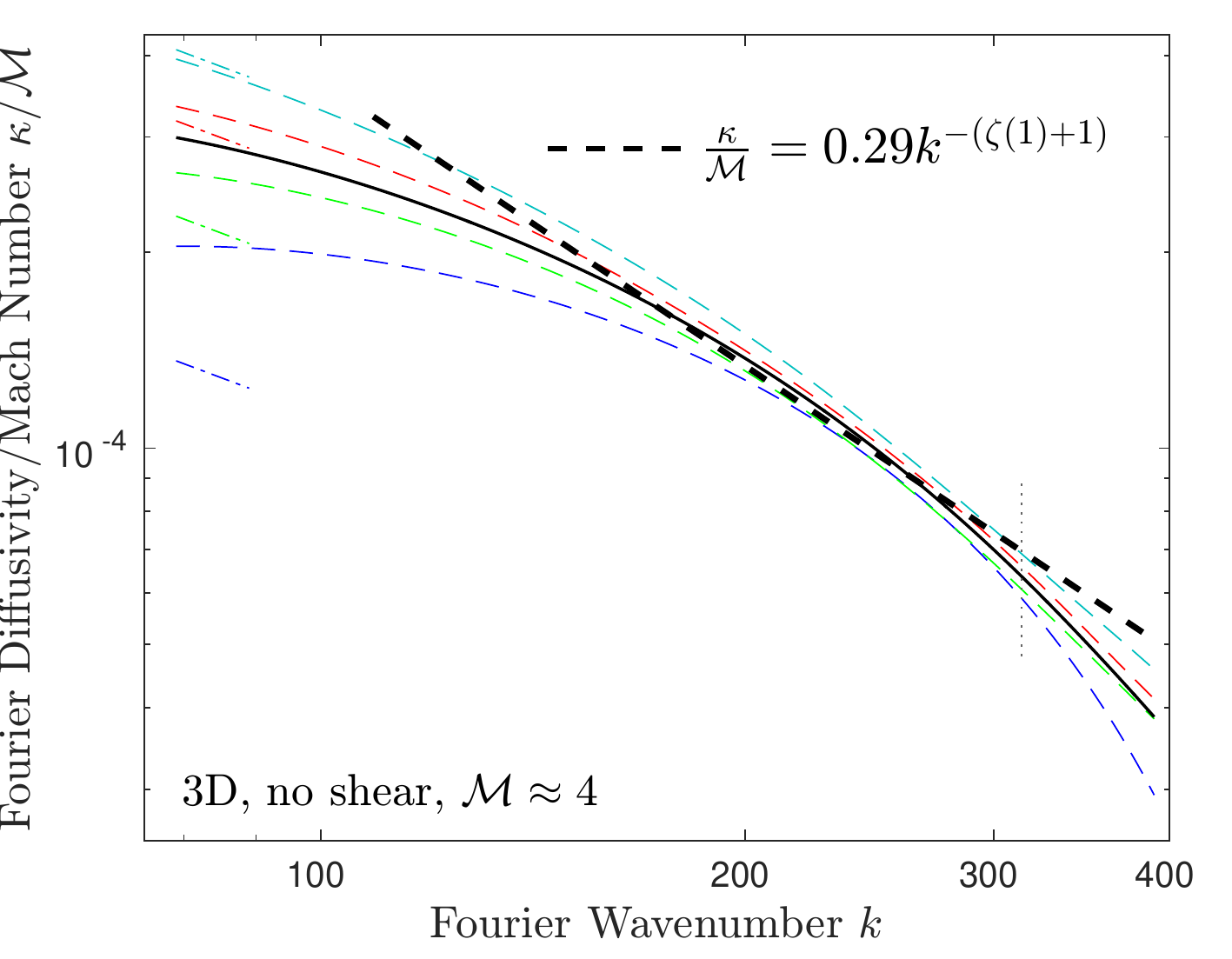} &

\includegraphics[width=0.97\columnwidth,trim={3.1cm 9cm 3.5cm 9.5cm},clip]{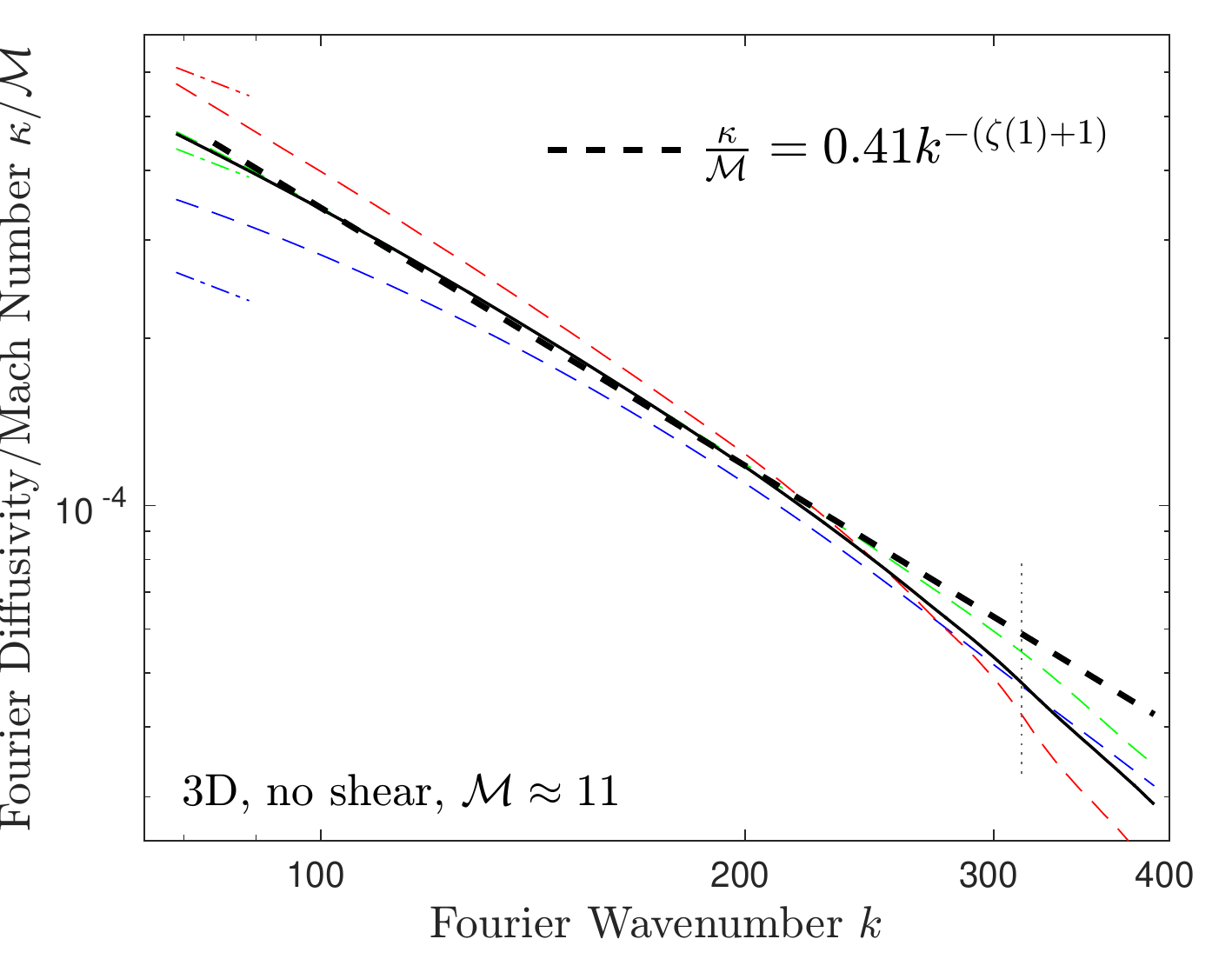} 
\end{tabular}
    \vspace{0.25cm}
    \caption{
    Same as Fig.~\ref{fig:kspaceHD}, but for each of the MHD simulations, as labeled at the bottom of each panel. In each simulation's inertial range (see Fig.~\ref{fig:struct}), we confirm a nearly time-independent scaling with universal constant coefficient $A\approx 0.3-0.6$, independent of Mach number, dimensionality, shear, and the presence of magnetic fields.
    \label{fig:kspace}}
\end{figure*}

\vspace{-0.5cm}
\section{Numerical Simulations}

\subsection{Code and Turbulent Driving}

The simulations here were run with {\sc gizmo}\footnote{A public version of the code, including all physics and numerical methods used in the simulations here, is available at:
\gizmourl} \citep{2015MNRAS.450...53H}, a mesh-free Lagrangian finite-volume Godunov code, in its Meshless-Finite Mass (MFM) mode. This method is designed to capture advantages of both smoothed-particle hydrodynamics (SPH) and grid-based adaptive-mesh refinement (AMR) methods. The advantages of the method are described and tested in extensive detail with a survey of $\sim 100$ test problems in \citet{2015MNRAS.450...53H,2015arXiv150502783H,hopkins:cg.mhd.gizmo}, for both HD and MHD, demonstrating good accuracy and agreement with well-studied regular-grid and moving-mesh Godunov codes. Of particular relevance to our studies here, these include both sub-sonic and super-sonic turbulence tests. 

In all cases considered in this paper, the turbulent driving routines, including parameters, follow \citet{2012MNRAS.423.2558B}. The usual box stirring method \citep*[e.g.][]{2008PhRvL.101s4505S,2008ApJ...688L..79F,2010MNRAS.406.1659P} is employed, with a small number of modes (wavelengths 1/2 to 1 times the box size) driven in Fourier space as an Ornstein-Uhlenbeck process in Fourier space with a mix of equal parts compressible and incompressible/solenoidal modes. For the case of our MHD simulations, we initialize a uniform seed field $\boldsymbol{\bf B}=B_0\,\hat{z}$; the seed field is chosen to have a trace initial value so that it is self-consistently amplified to saturation values by the turbulence (we do not consider cases with a strong mean-field such that the turbulence would be sub-Alfvenic). We discard all simulation outputs until all turbulent properties have reached a statistical steady-state (after the first few crossing times).

Our simulation with shear uses the standard shearing-sheet approximation \citep[see][]{2008ApJS..174..145G}. We solve the azimuthally-symmetric equations (following cylindrical $R$, $z$ coordinates) in a frame which co-rotates with circular orbits, with frame-centered orbital frequency $\Omega$. This amounts to adopting shear-periodic boundary conditions with centrifugal and Coriolis accelerations ${\bf a} = 2\,q\,x\,\Omega^{2}\,\hat{x} + 2\,{\bf v}\times(\Omega\,\hat{z})$ (where $q\equiv-d\ln{\Omega}/d\ln{R}=1$ here, for a constant-circular velocity disk).

\vspace{-0.5cm}
\subsection{Conventions and Units}

We briefly summarize conventions used. The wavenumber $k=|\boldsymbol{\bf k}|$ defines a ``length'' scale $\ell=2\pi/k\sim{}1/k$. We adopt an isothermal equation of state ($\gamma$ = 1) for the gas which is reasonable for the density and temperature ranges considered (given efficient cooling in the real ISM) and enables comparison with the ``standard'' ISM turbulence literature. Due to the scale-free nature of the fluid or MHD equations equations, in the fully-converged (infinite resolution) limit, the statistical properties of the turbulence are entirely determined by the dimensionless Mach number (the  mean magnetic field in the MHD runs is negligible compared to the turbulent velocities). We set the sound speed $c_{s}$, box length $L_{\rm box}$ and box mass $M_{\rm box}$ to unity in code units and define the Mach number $\mathcal{M}\equiv{{\langle{}v_t^2\rangle}^{1/2}}/{c_s}$, where ${\langle{}v_t^2\rangle}^{1/2}$ is measured at the box-scale.

\vspace{-0.5cm}
\subsection{Runs Performed}
\label{runs}


As our fiducial run, we first consider a 3D hydrodynamic simulation with $\mathcal{M} \approx 7$ with a resolution of $512^{3}$, testing the basic theory described in Sec.~\ref{theory1}.
To investigate the robustness of the ideas in physical situations with more realistic application to the ISM, we then  a variety of lower-resolution MHD runs: two 3D runs  at $\mathcal{M}\approx 4$ and $\mathcal{M}\approx 11$ and a resolution of $256^3$,  a 2D run at  $\mathcal{M}\approx 5$ (resolution $1024^2$), and a 2D case with a mean shear flow with $\Omega=10$ (at $\mathcal{M}\approx 2$, resolution $1024^2$), to account for rotation of a galactic disk. This value of $\Omega$ is chosen so the velocity scale-height $H\equiv \sqrt{c_{s}^{2} + v_{t}^{2}}/\Omega \approx 0.2$  (the turbulent driving scale is automatically set to $H$, appropriate for the driving scale being the disk scale-height in a stratified disk).
In the MHD simulations, since we initialize with a weak mean field, $\mathcal{M}$ also determines the saturated Alfv{\'e}n Mach number ($\mathcal{M}_A\equiv \langle v \rangle/\langle v_A \rangle$ where  $v_A=\left|\boldsymbol{\bf B}\right|/\sqrt[]{4\pi \rho}$ is the Alfv\'en speed). We $M_A$ consistent  with previous   studies (e.g., \citealt{kritsuk:2011.mhd.turb.comparison,federrath:supersonic.turb.dynamo}), varying from $\mathcal{M}_A \sim 1.5$ at lower $\mathcal{M}$ to $\mathcal{M}_A \sim 4$ at high $\mathcal{M}$.

\vspace{-0.5cm}
\subsection{Tracer Analysis}
\label{sec:tracer_release}

In this section, we describe our method for calculating  the diffusion properties of Lagrangian particles in the simulated turbulence. This method draws on the Lagrangian nature of the numerical method so as to not evolve separate equations for the scalar field.
Particle identities are stored throughout the simulation. Consider a single trial (denoted $\alpha$); in this trial each particle is assigned an initial scalar field $Z_n^{(\alpha)}$ ($Z$ for metallicity). We can then trace the particle through the simulation, reconstruct its transport, and repeat the ``injection'' in a new trial $\alpha^\prime$, with little computational effort. The tracer concentration (or metal density) for each trial, $\theta^{(\alpha)}$, can be determined via projection onto a fixed grid as 
\begin{eqnarray}
\theta_{ijk}^{(\alpha)}({\bf x},\,t)=\sum\limits_{{\rm particles}\,n} {{m_n}{Z_n^{(\alpha)}}}{W(\boldsymbol{{\bf x}}_{ijk}-\boldsymbol{{\bf x}}_{n}(t),h_{ijk})}
\label{density_field}
\end{eqnarray}
where $m_n$ is the mass of the nth particle, $i, j, k$ index the grid cells (with positions ${\bf x}_{ijk}$) in three dimensions, and $W$ denotes a cubic spline interpolant \citep[see][]{springel2011smoothed}, where $h_{ijk}$ is a spline kernel length adapted to enclose the nearest $\sim 64$ particles around each grid cell center (our results are not sensitive to this choice). The fixed grid for the projection is taken to be the same size as the simulation resolution (i.e. $512^3$ for the fiducial run, $256^3$ for the 3D MHD runs, and $1024^2$ for the 2D runs).
The initial metal density $\theta({\bf x},\,0)$, which determines $Z_n^{(\alpha)}$, is chosen to be a strongly peaked Gaussian with standard deviation 0.005 in code units. The choice of standard deviation  is chosen to allow a large number of times to be sampled before the size of the tracer cloud becomes comparable to the size of the box. It does not strongly affect the results discussed in Sec.~\ref{sec:results}.

We wish to capture the evolution of $\theta$ and work in a local frame where ${\mathbf{u}}=0$ to ignore simple bulk advection. Because the tracer is initially highly localized and we use periodic boundary conditions,  we may average over space by positioning the centre of the Gaussian at different points (for each trial) and taking the average of the tracer density at each time step. 
To sample the statistics of the saturated turbulent state, we
average over $N_{\rm trials}=1200$ tracer releases,  constructed by taking $15$ different initial injection times, each with $80$ different injection locations. For cases without shear the injection centers are randomly positioned anywhere in the box, whereas for the sheared case we sample from a plane tangential to the direction of background flow. At each time $t$, we center the grid on the center of the scalar field (which corrects for local advection and aims to capture the diffusion process in the Lagrangian frame of the mean velocity). We  then define the ensemble average: 
\begin{eqnarray}
\theta_{ijk}({\bf x},\,t)=\frac{1}{N_{\rm trials}}\sum\limits_{\alpha=1}^{N_{\rm trials}} \, {\theta^{(\alpha)}_{ijk}({\bf x},\,t)}
\label{density_field2}
\end{eqnarray}
Finally we average this in radial shells $r \equiv |{\bf x}|$, to obtain a radial profile $\theta(r,\,t)$. At long times after injection, when the profile/scalar distribution scale length becomes comparable to the box size, the periodic boundary conditions artificially corrupt further evolution, so we consider only those times before the profile has been distorted by the edge of the box.

To calculate $\widehat{{\theta}}$ we numerically compute
\begin{eqnarray}
\widehat{\theta}(\boldsymbol{\bf k},t)=\int\limits_{\Re^n} \theta(r,\,t)\exp(-i\,\boldsymbol{\bf r}\boldsymbol{\cdot}\boldsymbol{\bf k})\ \mathrm{d}\boldsymbol{\bf r}\,,
\end{eqnarray}
using the radially averaged $\theta(r,\,t)$. 
We then approximate the time-derivative using a finite-difference 
\begin{eqnarray}
{\partial_t}\widehat{{\theta}}\approx\frac{\widehat{{\theta}}(k,t+\Delta{}t)-\widehat{{\theta}}(k,t-\Delta{}t)}{2\Delta{}t}
\end{eqnarray}
for $\Delta{}t=0.005$, to allow computation of the diffusivity Eq.~\eqref{kappadef2}.

\vspace{-0.5cm}
\section{Results}\label{sec:results}

\subsection{Scalar Density Distributions: Simple, Constant Diffusivity Cannot Describe the Simulations}
\label{sec:diffusivity.not.constant}

As an illustrative example of our passive-scalar tracing procedure, Fig. \ref{fig:averaging} shows the evolution of one tracer injection/release, and compares it to the ensemble average over $\sim1200$ tracer releases randomly distributed in the turbulence, as described above. Both of these are taken from our fiducial run.  One can see that for an {\em individual} injection, after a time $\sim 0.03$ (a few eddy turnover times for eddies with length of order the initial injection spatial scale), the dominant tracer motion is simple advection, with some significant distortion of the (initially Gaussian) profile. The distribution does not, in any meaningful sense, resemble the solution to a diffusion equation. However, when we {\em ensemble-average}, the distribution shows behavior much more similar to our expectations for diffusive processes. The distribution is approximately radially symmetric about the injection site -- i.e.\ there is no ensemble-average preferred direction (this is also true for the MHD runs) and the distribution falls off radially in diffusion-like manner.  This behavior is simply a consequence of averaging over a large number of realizations after removing the mean advective motion. The same averaging effect is used in all other simulations, with the same results.

Figure~\ref{fig:tails} shows the ensemble-averaged scalar density profiles $\theta(r,\,t)$ for the 2D unsheared simulation. 
Although the initial profile of the injected tracer is, by construction, Gaussian, the profile develops thicker, highly non-Gaussian ``superdiffusive'' tails rapidly. If the ``effective diffusivity'' were scale-independent ($\kappa = \mathrm{const}$) then Eq.~\eqref{equation_of_state} would give the standard Gaussian diffusion solution: $\theta(r,\,t) \propto \exp{[-r^{2}/4\,t\,\kappa]}$. 
However, as shown in  Sec.~\ref{theory1} (Eq.~\ref{FTconvention2}), if $\kappa\propto{}k^{-\alpha}$, then $\theta(r,\,t) \propto t\,r^{-(2+n-\alpha)}$ where $n$ is the dimension of the system and $\alpha\approx 3/2$  for supersonic turbulence. This gives $\theta(r,\,t)\propto r^{-5/2}$ in 2D and $\propto r^{-7/2}$ in 3D, agreeing very well with the measured profile in Fig.~\ref{fig:tails}.
 This  illustrates that the process is well described by a fractional diffusion with the scaling expected from supersonic turbulence.  The 3D runs show  similar behavior, but due to the somewhat lower resolution, the late-time power-law behavior is less well defined.  
In particular, by the time the tails develop the asymptotic scaling, the tracer has begun to diffuse to the boundary of our box, illustrating the main difficulty in accurately measuring $\kappa$ in simulation.

\vspace{-0.5cm}
\subsection{Fourier Scalings: How Does Diffusivity Depend on Scale?}\label{sec:Fourier scaling results}

For comparison with the simple theory outlined in Sec.~\ref{theory1} we must first estimate the structure functions of the gas velocity distributions.  We do this by selecting a sample size of $10^{10}$ random particle pairs and calculating the average $S_p(l) = \left\langle{{\left|{\boldsymbol{\bf v}}({\boldsymbol{\bf x}}+{\boldsymbol{\bf l}})-{\boldsymbol{\bf v}}({\boldsymbol{\bf x}})\right|}^p}\right\rangle$ (Eq.~\ref{structure}) over such samples at all times considered in the simulation. Fig.~\ref{fig:struct} shows the structure functions for $p=1,2$ plotted against $l$ (in codes units) and fit with power laws to give an estimate for $\zeta(p)$ in each simulation. We fit between $l>l_{\mathrm{fit}} = 2\times 10^{-2}$ and the largest scales (where $S_p(l)$ is flattened due to the influence of turbulent driving). This range is chosen in the fiducial simulation to be above the point at which the velocities become transonic ($v_l\sim c_s$), which is where  $S_p(l)$ flattens (this effect is evident in the top panel of Fig.~\ref{fig:struct}).  For simplicity, 
we use the same $l_{\mathrm{fit}}$ in the other simulations (bottom panel of Fig.~\ref{fig:struct}), since $S_p(l)$ in each is approximately a power law for $l>l_{\mathrm{fit}}$ and the exact choice of $l_{\mathrm{fit}}$ does not make a significant difference to the measured $\zeta(p)$.\footnote{In the 2D shearing simulation, which is at lower Mach number, the 
velocities are subsonic at $l\sim l_{\mathrm{fit}}$. However, the measured $S_p(l)$ (see Fig.~\ref{fig:struct}) are close to power law anyway.} For the 2D run with shear, we measure the structure functions with ${{\bf l}}$ transverse to the mean-flow direction. In general, the exponents $\zeta(p)$ are not universal and depend on Mach number and how the turbulence is driven. Our measured values are comparable to those seen in previous literature \citep{Lee2003-2Dturb,kritsuk2007statistics,schmidt2009numerical,kritsuk2013supersonic}, increasing somewhat with $\mathcal{M}$ without any significant differences between 2D or 3D domains.

We now calculate the ``Fourier Diffusivity'' (effective diffusivity associated with modes of wavenumber $k$), $\kappa=-{\partial_t}\widehat{{\theta}}/({k^2}\widehat{{\theta}})$, as described in \S~\ref{theory1}. 
This is done as a function of $k$, at different times $t$. In each case, we use the values of $\zeta(n)$ quoted in Fig.~\ref{fig:struct} to compare to the measurements of $\kappa$.

Results for the fiducial simulation shown in shown in Fig.~\ref{fig:kspaceHD}. There is a  region at moderate $k$ where the  diffusivities approximately coincide and are independent of time. We overplot the expected power-law scalings from our theory and the measured structure function scaling $\zeta(1)\approx 0.43$ (see Eq.~\ref{Dscale2}). The agreement between the theory and measurements is seen to be good. In particular, we see decent agreement with the expected scaling $\kappa \sim A\,\mathcal{M}\,k^{-(\zeta(1)+1)}$ for an intermediate range of scales at moderate $k$.  Deviations from the scaling at the largest $k$ owe to (1) limited numerical resolution, and (2) reaching the sonic scale where the turbulence becomes sub-sonic (this occurs at $k\sim k_{\mathrm{fit}}=2\pi/l_{\mathrm{fit}}$, shown with dotted line; see Fig.~\ref{fig:struct}).  At small $k$, we transition to ``ballistic motion'' dominating the transport and see qualitative agreement with expected scaling $\kappa \sim t\,k^{-\zeta(2)}$. In particular, we observe a flattening of $\kappa$, which moves to larger scales in time, and the normalization increases with time as expected (although the scaling is somewhat slower than linear in $t$). 

We now consider the same analysis for the MHD simulations across a wider range of Mach numbers, with the results for each simulation shown in Fig.~\ref{fig:kspace}. Again we see a scaling range where the time-independent power law agrees  with Eq.~\eqref{expected_rel} (with $\zeta(1)$ taken from the measurements in Fig.~\ref{fig:struct}). We also see a similar normalization,  $A\approx0.3-0.6$, in all simulations, independent of Mach number, dimensionality, and shear. Again, there are we  are deviations  from the power-law scaling at large $k$, which is more severe in the 3D MHD runs due to the lower resolution ($256^3$). We also see reasonable agreement with the ballistic motion prediction, $\kappa \sim t\,k^{-\zeta(2)}$, for very low $k$, although the linear dependence on time overpredicts the measured increase in $\kappa$ (as in Fig.~\ref{fig:kspaceHD}, but this seems particularly true at lower $\mathcal{M}$). Note that for the $\mathcal{M}\sim 11$ simulation, we show a reduced number of times because the increased Mach number leads to the pollution of the measurement by the box boundary at late times. 

Comparing all runs, we see similar  qualitative features and generic  agreement with the scalings outlined in Sec.~\ref{theory1}. Most importantly, we see the expected steepening of $\kappa$ in the inertial range for the larger-$\mathcal{M}$ simulations where $\zeta(1)$ is larger. In addition, as noted above, the normalization parameter $A$ is consistent across all simulations, and the temporal change in $\kappa$ at low $k$ is qualitatively consistent with the model (although somewhat slower than linear in time).  The independence on the dimensionality and  the details of the gas physics -- for example, the presence of magnetic fields or a mean shear flow -- is also expected, since supersonic turbulence is dominated by the strong shocks, and the differences between 2D and 3D are less extreme than for subsonic turbulence. Similarly, the magnetic field, being less efficiently amplified in supersonic turbulence compared to subsonic turbulence \citep{federrath:supersonic.turb.dynamo}, plays a subsidiary role (so long as  $\mathcal{M}_A >1 $, otherwise the turbulence will be more Alfv\'enic in character; see \citealt{2002PhRvL..88x5001C}). In the shearing case, we also see similar results. At large $k$ this may be expected, since the shear velocity $\Delta v_{\rm shear} \sim \Omega \ell$ is sub-dominant to the turbulent velocities below the velocity scale-length $H= (c_{s}^{2} + v_{t}^{2})^{1/2}/\Omega \approx 0.2$. By the time the diffusion expands beyond these scales, it is directly  influenced by the boundary conditions (but in any case, the shearing-box approximation is no longer valid on scales $\gg H$). So we caution that our ``with shear'' results are {\em not} necessarily valid in the regime where shear {\em dominates} the motion, but only when it is present but secondary to turbulence. Nonetheless, it is significant on the small scales -- it causes an ``aliasing'' (a slight elliptical distortion of the tracer cloud) if we do not properly account for it in the analysis (this is done by transforming the $y$ coordinate to $Y=y+Stx$, which factors out the linearized ``pure shear'' motion on our initial tracer injection, leaving the truly diffusive component).

\vspace{-0.5cm}
\section{Conclusions}

We suggest simple scaling arguments for the diffusion of passive scalars in supersonic turbulence, based on Richardson diffusion with modified velocity scalings \citep{Richardson709}. These ideas are then tested on a variety of numerical simulations of neutral and MHD supersonic turbulence in two and three dimensions. We summarize our conclusions as follows:

\begin{enumerate}
	\item We show that the ``effective diffusivity'' $\kappa$ cannot be constant, i.e. the scalar density does not obey a pure diffusion equation $\partial_{t}Z = \kappa\nabla^{2}Z$, with $\kappa=$\,constant. This would conserve a Gaussian-like profile; however, if we inject tracers to follow their evolution, we see large non-Gaussian tails appear immediately, indicating that $\kappa$ must be scale-dependent.
	\item We demonstrate the existence of an effective time-independent scale-dependent diffusivity $\kappa(k)$, which explains the non-Gaussian features and time-dependence described above, and is invariant over a suitable range of scales $\ell=2\pi/k$ (corresponding to the inertial range). This scaling is approximately given by:
\begin{eqnarray}
\kappa\approx 0.5\mathcal{M}_{L_{\rm box}}\,(k\,L_{\rm box})^{-\alpha}
\end{eqnarray}
where $\mathcal{M}_{L_{\rm box}} = \mathcal{M}(L_{\rm box}) = \mathcal{M}$ is the Mach number (defined at some normalization scale, here the box scale $L_{\rm box}$) and the scaling exponent $\alpha$ increases weakly from $\approx 1.45$ at Mach numbers $\mathcal{M}\approx{}4$ to $\approx 1.54$ at Mach numbers $\mathcal{M}\approx{}11$. In other words, the system can be modeled as a diffusion process, but with each mode of the tracer density field obeying a separate diffusion equation according to its mode-dependent diffusivity.
	\item The exponents $\alpha$ agree well with arguments based on the velocity structure functions of the turbulence in the inertial range. Dimensionally, if eddies of scale $\ell$ advect or mix material on their crossing time, we expect a scaling of $\kappa \sim v_{t}(\ell)\ell$, where $v_{t}(\ell) \propto \ell^{\zeta(1)}$ is the characteristic eddy velocity on scales $\ell$. Based on the phenomenology of supersonic turbulence, we expect $\zeta(1)\sim0.5$, possibly increasing with Mach number (see \citet{2007AIPC..932..421K,2007AAS...21113803K,2008PhRvL.101s4505S,2010A&A...512A..81F,2010MNRAS.406.1659P,2010PhLA..374.1039S}), giving $\kappa\sim \ell^{1.5}$. This is almost exactly what we measure directly, including the weak dependence on Mach number, which indicates the validity of the simple phenomenological arguments.
	\item We identify a superdiffusive regime at large scales and small times, where the eddies simply transport the particles via bulk advection (``ballistic motion''), leading to the scaling
	\begin{eqnarray}
	\kappa\propto t\,\ell^{\zeta(2)}.
	\end{eqnarray}
	\item We demonstrate that these statements are only valid in a statistical, {\em ensemble-averaged} sense: any individual Lagrangian parcel of fluid can be distorted into a high non-symmetric shape which bears no resemblance to the isotropic solution of a diffusion equation, and can remain coherent for many turbulent crossing times. Diffusive behavior only appears after ensemble-averaging over all possible behaviors. This has important implications for physical systems: for e.g.\ metal mixing, it means that ``metal diffusion'' only applies when the number of ``sources'' is large and well-distributed in time and/or space. If we consider the material injected by e.g.\ just a single SNe explosion (or, on larger scales, the SNe from a single star cluster), which may be very important for the second-generation of star formation, this material does not simply diffuse but may create long-lived ``pockets'' of enriched gas. 
	\item We show that our scalings above remain true in the presence of a coherent shear force -- at least on scales where the shear velocity is sub-dominant to the turbulent motions -- once the simple shear has been accounted for in the tracer profile. Clearly, more study of the shearing case is warranted to develop fundamentally anisotropic scalings that can be applied even in the regime where shear motions are larger than turbulent motions.
	\item We do not see strong effects, either in the ensemble-averaged statistical anisotropy, or scaling exponents, from magnetic fields. This is consistent with the fact that (in 3D) the saturation of the supersonic turbulent dynamo produces super-Alfvenic turbulence. However, we caution that the imposition of a sufficiently strong mean magnetic field (strong enough to make the turbulence sub-Alfvenic) will likely lead to different results.
	\item For purposes of subgrid-scale models, our scalings imply that the ``effective turbulent diffusivity'' is not, in fact, a constant. However, if $\kappa \propto v_{t}(\ell)\ell$ for modes of wavelength $\ell$, then sufficiently large wavelength modes $\ell \gg \Delta$ (where $\Delta$ is the simulation grid-scale) will always have their mixing resolved. Meanwhile since $v_{t}(\ell) \propto \ell^{\beta}$ with $\beta\sim 0.5>0$, the effect of unresolved, small-scale modes will be {\em dominated} by the largest un-resolved modes, i.e. those with $\ell \sim \Delta$. Therefore a scaling of the form typically adopted in Smagorinsky models is formally justified by our analysis, {\em provided} the following conditions are met: (a) the scale $\Delta$ lies {\em within the inertial range} of the turbulence, (b) the velocity components identified by the shear tensor $S$ are genuinely turbulent, and not some other (gravitational, outflow, inflow) motion, (c) the turbulence is statistically isotropic, and (d) shear is negligible on the scale $\Delta$, as given by our note (vi) above. We stress that if any of these conditions is violated, the sense of the error will generally be that the Smagorinsky prescription {\em over-estimates} the diffusivity, potentially by very large factors. Moreover, we also emphasize that the constant pre-factor in such scalings must be calibrated to the appropriate definition of the grid scale $\Delta$ -- this must be done independently for different numerical methods, because they have different ``effective resolution scales'' of the turbulent cascade, so we do not quote an effective value for it here.
\end{enumerate}

We have focused on a simple, limited set of simulations illustrating some of the key turbulent processes controlling the diffusion of metals and other passive scalars in the ISM. Of course, more detailed physical simulations including realistic phase structure and mixing by non-turbulent processes (e.g.\ galactic winds and fountains) will be necessary for a complete picture of mixing in realistic physical systems.

\vspace{-0.7cm}
\acknowledgments 
Support for MJC was provided by the SURF program at Caltech and St John's College, Cambridge. 
Support for PFH \&\ XM was provided by an Alfred P. Sloan Research Fellowship, NASA ATP Grant NNX14AH35G, and NSF Collaborative Research Grant \#1411920 and CAREER grant \#1455342. Support for JS was provided by the Sherman Fairchild foundation, and by the Gordon and Betty Moore Foundation
through Grant GBMF5076 to Lars Bildsten, Eliot Quataert and E. Sterl
Phinney. Numerical calculations were run on the Caltech compute cluster ``Zwicky'' (NSF MRI award \#PHY-0960291) and allocation TG-AST130039 granted by the Extreme Science and Engineering Discovery Environment (XSEDE) supported by the NSF. \\

\vspace{-0.2cm}
\bibliography{./ref}

\end{document}